\documentclass[10pt]{article}

\begin{document}

\author{Angelo Tartaglia, Matteo Luca Ruggiero \\
Dip. Fisica, Politecnico, and INFN Torino, Italy \\
angelo.tartaglia@polito.it, matteo.ruggiero@polito.it}
\title{Gravitomagnetic measurement of the angular momentum of celestial
bodies}
\maketitle

\begin{abstract}
The asymmetry in the time delay for light rays propagating on
opposite sides of a spinning body is analyzed. A frequency shift
in the perceived signals is found. A practical procedure is
proposed for evidencing the asymmetry, allowing for a measurement
of the specific angular momentum of the rotating mass. Orders of
magnitude are considered and discussed.
\end{abstract}

\vspace{2cm}

A well known effect of gravity on the propagation of
electromagnetic signals is the time delay: for example an
electromagnetic beam emitted from a source on Earth toward another
planet of the solar system, and hence reflected back, undergoes a
time delay during its trip (with respect to the propagation in
flat space-time), due to the influence of the gravitational field
of the Sun. This effect was indeed called the "fourth" test of
General Relativity\cite{shapiro64},\cite{shapiro66}, coming fourth
after the three classical ones predicted by
Einstein\cite{einstein}.

Though small, the time delay in the propagation of electromagnetic
waves was detected by Shapiro et al. \cite{shapiro}, timing radar
echoes from Mercury and Venus, by means of the radio-telescopes of
Arecibo and Haystack. Anderson et al. \cite{ander} measured the
time delay of the signals transmitted by Mariner 6 and 7 orbiting
around the Sun. Finally Shapiro and Reasenberg obtained more
accurate results using a Viking mission that deposited a
transponder on the surface of Mars: the theoretical prediction was
verified within $\pm 0.1\%$ \cite{marte1} \cite{marte2}.

These measurements accounted just for the presence of a massive
source, described by the Schwarzschild solution of the Einstein
field equations. However, there is another correction to the time
delay, due to the spin of the source, which is produced by the so
called "gravitomagnetic" interaction\cite{ruggierotartaglia02}.

The time delay in the gravitational field of the Sun can be
studied in the weak field approximation\cite{straumann}. By using
Cartesian coordinates and assuming that the $z$ axis coincides
with the direction of the angular momentum of the source, the
propagation of electromagnetic signals, in the equatorial plane,
is described by the null world-line
\begin{equation}
0=g_{tt}dt^{2}+g_{xx}dx^{2}+g_{yy}dy^{2}+2g_{xt}dxdt+2g_{yt}dydt
\label{eq:metrica1}
\end{equation}%
In the weak field approximation, the bending of the trajectories
due to the gravitational field is neglected. This assumption,
appropriately choosing the $x$ and $y$ axes, leads to a ray
trajectory that is a straight line $x=b=const$; $b$ is of course
the closest approach distance with respect to the spinning body.
On using the appropriate form of the metric
elements\cite{tartaglia04}, the time of flight of the
electromagnetic beams can be written as
\begin{equation}
t_{f}(y_{1},y_{2})=t_{0}+t_{M}+t_{J}  \label{eq:delayc}
\end{equation}%
where
\begin{eqnarray}
t_{0} &=&\frac{y_{2}-y_{1}}{c}  \label{eq:tzero} \\
t_{M} &=&\frac{2GM}{c^{3}}\ln \frac{y_{2}+\sqrt{b^{2}+y_{2}^{2}}}{y_{1}+%
\sqrt{b^{2}+y_{1}^{2}}}  \label{eq:temme} \\
t_{J} &=&\mp \frac{2GJ}{c^{4}b}\left[ \frac{y_{2}}{\sqrt{b^{2}+y_{2}^{2}}}-%
\frac{y_{1}}{\sqrt{b^{2}+y_{1}^{2}}}\right]   \label{eq:temmea}
\end{eqnarray}%
In Eqs. (\ref{eq:temme}) and (\ref{eq:temmea}) $M$ is the mass of
the source and $J=Mca$ is its angular momentum, assumed to be
orthogonal to the plane of the motion.

The quantity $y_{1}$ is the $y$ coordinate of the source of the signals, $%
y_{2}$ is the $y$ coordinate of the receiver.  The time $%
t_{0}$ is clearly the Newtonian time of flight; $t_{M}$ is the
gravitational time delay measured by Shapiro \textit{et al.}, and
$t_{J}$ is the correction to the time delay produced by the
gravitomagnetic interaction with the angular momentum of the
central body. The double sign in $t_{J}$ means that
gravitomagnetism shortens the time of flight on the left and
lengthens it on the right.

This asymmetry, within the solar system, is in any case small:
however its systematicness lends an opportunity to reveal the
effect, appropriately combining the ticks of a 'clock' passing
behind the spinning mass. Elsewhere\cite{tartaglia04} we have
outlined a way to reveal this effect, based on the fact that the
relative motion of source, receiver and central mass, produces a
varying time delay, which shows up as a small frequency shift.
The gravitomagnetic contribution to this shift is manifested as an
asymmetry between right and left with respect to the central mass.
Let  $\nu _{0}$ be the proper frequency of the electromagnetic
signal; in these conditions, mirroring the signal after the
occultation (time reflection), then superposing corresponding
records with the signal before the occultation of the clock, a
beating function will result, where the magnitude of the frequency
of the beats is proportional to the angular momentum of the
spinning body
\begin{equation}
\nu _{2}=4\nu _{0}\frac{\mu a}{b^{2}}\frac{v_{0}}{c}
\label{modula}
\end{equation}%
and the frequency of the basic signal is shifted with respect to
the flat space-time situation by an amount proportional to the
mass:
\begin{equation}
\nu _{1}=\nu _{0}\left( 1+4\frac{\mu }{b}\frac{v_{0}}{c}-\frac{\mu b}{R^{2}}%
\allowbreak \frac{v_{0}}{c}\right)   \label{portante}
\end{equation}%
In Eqs. (\ref{modula}) and (\ref{portante}) $v_{0}$ is the
apparent transverse velocity of the source in the sky, $R$ is the
distance of the observer from the central body and $\mu
=G\frac{M}{c^{2}}$. To fix a few numbers, let us consider the
situation in the solar system, with the Sun as the central
spinning body, and an Earth bound observer. The source is a far
away astronomical body (e.g. a pulsar). In this case the orders of
magnitude are
\[
\begin{array}{ccccc}
\mu \sim 10^{3}\mathit{\ m} & a\sim 10^{3}\mathit{\ m} & b\sim 10^{9}\mathit{%
\ m} & R\sim 10^{11}\mathit{\ m} & v_{0}\sim 10^{4}\mathit{\ m/s}%
\end{array}%
\]
As a consequence we have
\begin{eqnarray}
4\frac{\mu }{b}\frac{v_{0}}{c} &\sim &10^{-10}  \nonumber \\
\frac{\mu b}{R^{2}}\allowbreak \frac{v_{0}}{c} &\sim &10^{-14}
\label{stime}
\\
4\frac{\mu a}{b^{2}}\frac{v_{0}}{c} &\sim &10^{-16}  \nonumber
\end{eqnarray}
The frequency effect, connected with a varying time delay in the
propagation of electromagnetic signals in the vicinity of a
spinning massive body, is very small, however not entirely
negligible, at least when a suitable procedure to detect it is
used, such as the one we have just outlined, based on the
possibility of producing a beat between signals after and before
the occultation.

Of course many practical problems have to be considered and
discussed to transform some principle formulae into an actual
measurement. Other situations of physical interest can be taken
into account: for instance, a more favorable observational
condition (with regard to the amount of the effect) can be the one
of a pulsar orbiting a neutron star. For more details, see the
discussion in our paper\cite{tartaglia04}. What we can say, at
this stage, is that the method we propose seems promising for
detecting gravitomagnetic effects on the propagation of
electromagnetic signals, and for measuring the angular momenta of
celestial bodies.

\end{document}